\documentclass[aps,prb,twocolumn,showpacs,preprintnumbers,amsmath,amssymb,superscriptaddress,longbibliography,nofootinbib]{revtex4-2}
\usepackage{hyperref}
\hypersetup{colorlinks=true, citecolor=blue, urlcolor=blue, linkcolor=blue,urlcolor=cyan}
\usepackage{color}
\usepackage{graphicx}
\usepackage{dcolumn}
\usepackage{braket}
\usepackage{amsthm}
\usepackage[caption=false]{subfig}

\newcommand{\ex}[1]{\mathrm{e}^{#1}}
\newcommand{\fr}{\frac}

\newcommand{\ca}[1]{\mathcal{#1}}
\newcommand{\bb}[1]{\mathbb{#1}}

\def\del{{\partial}}

\makeatletter
\let\cat@comma@active\@empty
\makeatother

\begin{document}
\preprint{KYUSHU-HET-341}
\preprint{RIKEN-iTHEMS-Report-25}
\title{Transmission Coefficients from Phantom Currents}

\author{Yuma Furuta }\email[]{furuta.yuma@phys.kyushu-u.ac.jp}
\affiliation{\it Institute for Advanced Study, 
Kyushu University, Fukuoka 819-0395, Japan.}

\author{Yuya Kusuki}\email[]{kusuki.yuya@phys.kyushu-u.ac.jp}
\affiliation{\it Institute for Advanced Study, 
Kyushu University, Fukuoka 819-0395, Japan.}
\affiliation{\it Department of Physics, 
Kyushu University, Fukuoka 819-0395, Japan.}
\affiliation{\it RIKEN Interdisciplinary Theoretical and Mathematical Sciences (iTHEMS),
Wako, Saitama 351-0198, Japan.}

\author{Toshiki Onagi}\email[]{toshiki.onagi@yukawa.kyoto-u.ac.jp}
\affiliation{\it Yukawa Institute for Theoretical Physics, Kyoto University, Kitashirakawa Oiwakecho, Sakyo-ku, Kyoto 606-8502, Japan.}

\begin{abstract}
A representative quantity that characterizes the dynamics of conformal interfaces is the transmission coefficient, 
which is defined through correlation functions of the stress tensor.  
Typically, this coefficient is complicated and highly dependent on its details.
In this work, we introduce a new perspective based on the notion of a ``phantom current''.  
We have shown that a spin-2 phantom current arising from the folding trick completely determines the transmission coefficient.  
In particular, when there is a single phantom current, the transmission coefficient is uniquely fixed by its conformal dimension.  
As a result, our framework provides a unified explanation of known results in minimal models and the free boson, 
while also yielding concrete predictions for previously unexplored interfaces.
\end{abstract}
\maketitle

\section{Introduction and Summary}

An {\it interface} is a codimension-one defect that connects two (possibly different) quantum field theories.
Interfaces give rise to a wide range of phenomena and have been actively studied from various perspectives, such as energy and charge transmission or reflection \cite{Quella2007,Brunner2015,Meineri2020,Bachas2020,Bachas_2023,Karch2024,Liu2025},
as well as entanglement and information propagation \cite{Sakai2008,Calabrese2012,Brehm2015,Wen2018,Karch2023,Tang:2023chv,Karch2024,Baig2024,Barad2025,Afxonidis2025}.  
More recently, it has been recognized that (not necessarily group-like) topological interfaces can play a role analogous to symmetries \cite{Gaiotto2014}, 
and they have been extensively discussed in the context of generalized symmetries \cite{Shao2023}.
Conformal field theory (CFT), in particular, provides a controlled laboratory in which all these aspects can be analyzed.  
Furthermore, in the AdS/CFT correspondence, interfaces correspond to branes on the quantum gravity side \cite{Azeyanagi2008,Takayanagi2011,Fujita2011},
offering a powerful framework for non-perturbative analyses of quantum gravity involving branes.

In 1+1--dimensional CFTs, a conformal interface is defined as an interface that preserves the diagonal Virasoro symmetry \cite{Oshikawa1996,Oshikawa1997,Bachas2002},
which translates into the following boundary condition on the stress tensor:
\begin{equation}
T_L-\overline{T}_L=T_R-\overline{T}_R,
\end{equation}
where $T_L$ and $T_R$ denote the holomorphic stress tensors on the left and right sides of the interface, respectively, while $\overline{T}_L$ and $\overline{T}_R$ denote the corresponding anti-holomorphic ones.
Physically, this corresponds to imposing the energy conservation law, which states that no energy is absorbed or emitted at the interface.
A central quantitative measure of interfaces is the {\it transmission coefficient}, which encodes how much energy is transmitted across the interface \cite{Quella2007}.
The transmission coefficient is controlled by the two-point function of the stress tensors across the interface \cite{Billo2016, Meineri2020},
\begin{equation}
\braket{T_L(z)T_R(w)} = \fr{c_{LR}}{2(z-w)^4}+\cdots.
\end{equation}
The weighted average transmission coefficient defined in \cite{Quella2007} can be expressed in terms of $c_{LR}$ as
\begin{equation}
\ca{T} = \fr{2c_{LR}}{c_L+c_R}, \label{eq:defclr}
\end{equation}
where $c_L$ and $c_R$ represent the central charges of the left and right CFTs, respectively.
Typically, the transmission coefficient depends intricately on the details of the interface.
For example, the transmission coefficient is generally independent of the $g$--function, introduced in \cite{Affleck1991}.
Furthermore, the transmission coefficient is also independent of the effective central charge, which measures the amount of information transmitted \cite{Karch2023,Karch2024}.

\begin{figure}[t]
 \begin{center}
  \includegraphics[width=7.0cm,clip]{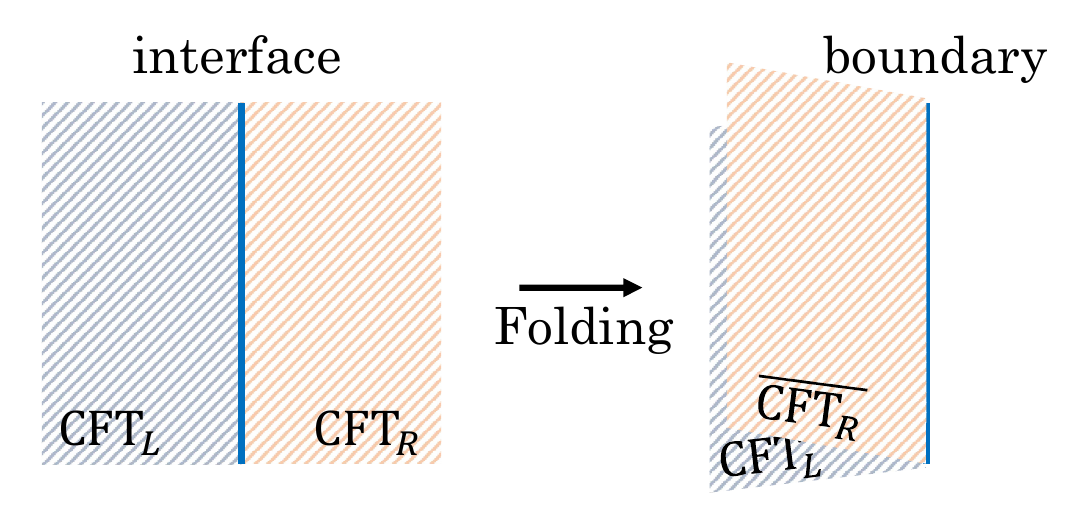}
 \end{center}
 \caption{The folding trick enables the reinterpretation of an interface between $\mathrm{CFT}_L$ and $\mathrm{CFT}_R$ as a boundary in $\mathrm{CFT}_L \otimes \overline{\mathrm{CFT}_R}$.
}
 \label{fig:folding}
\end{figure}

This Letter develops a general framework that predicts the transmission coefficient from {\it minimal} input data.
Motivated by the Gaiotto renormalization group (RG) brane \cite{Gaiotto2012, Poghosyan2014}, we focus on theories whose folded description contains only two spin-2 currents -- the stress tensor and a {\it spin-2 phantom current}.
The phantom current refers to a hidden current that does not exist in either $\mathrm{CFT}_L$ or $\mathrm{CFT}_R$ individually,  
but emerges in the tensor product CFT obtained via the {\it folding trick} \cite{Wong1993,Oshikawa1996,Oshikawa1997} (see Fig.~\ref{fig:folding}).
The spin-1 phantom current explains how a CFT without continuous symmetries can admit a defect conformal manifold \cite{Antinucci2025}.
Complementarily, this article elucidates the role played by the spin-2 phantom current in interface CFTs.

We demonstrate that the transmission coefficient is completely determined by the conformal dimension of the phantom current alone.
Our framework explains disparate known results within a single mechanism,
including the transmission coefficient in minimal models \cite{Brunner2015} and free bosons \cite{Quella2007}.
Beyond benchmarks, we obtain concrete predictions for interfaces whose explicit construction is not known, e.g. the RG brane for $\mathcal{M}(2q{+}1,q)\to\mathcal{M}(q,2q{-}1)$,
providing testable targets for numerics in non-unitary CFTs.

\section{Phantom Currents in Minimal Models}

\newsavebox{\boxcomp}
\sbox{\boxcomp}{\includegraphics[width=7cm]{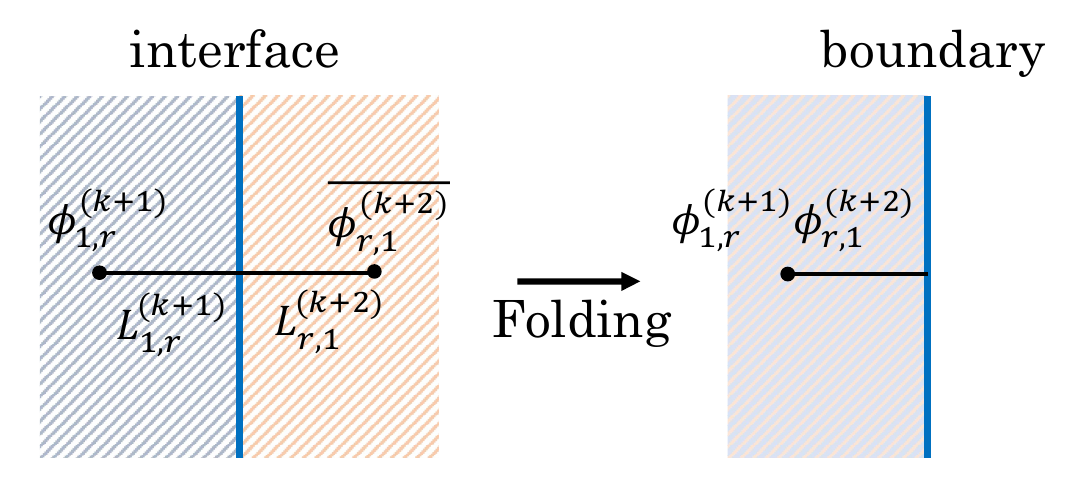}}
\newlength{\compw}
\settowidth{\compw}{\usebox{\boxcomp}}

It is extremely rare to find examples of non-topological interfaces whose exact constructions are known.
Among the few known examples, one is the {\it Gaiotto RG brane} \cite{Gaiotto2012, Poghosyan2014},
which is an interface between two nearby minimal models, $\mathcal{M}(q+1,q)$ and $\mathcal{M}(q,q-1)$, connected by the {\it Zamolodchikov RG flow} \cite{Zamolodchikov1987}.

The minimal model $\mathcal{M}(k+3, k+2)$ is a rational conformal field theory (RCFT) based on the Virasoro algebra,  
characterized by the following central charge
\begin{equation}
c^{(k+2)}=1-\fr{6}{(k+2)(k+3)}
\end{equation}
and spectrum
\begin{equation}
h^{(k+2)}_{r,s}=\fr{(r-s)^2}{4}+\fr{r^2-1}{4(k+2)}-\fr{s^2-1}{4(k+3)}.
\end{equation}
The allowed values of the Kac labels $(r, s)$ are given by $1 \leq r \leq k + 2$ and $1 \leq s \leq k + 3$.
Minimal models may allow non-diagonal modular invariants,  
but in what follows, we consider only the diagonal modular invariant.

Minimal models are known to exhibit non-invertible symmetries generated by Verlinde lines $\{ L^{(k+2)}_{r,s}\}$\cite{Verlinde1988}, which act on primary states (i.e. $\mathrm{Vir}\otimes \overline{\mathrm{Vir}}$ highest weight states) as
\begin{equation}
L^{(k+2)}_{r,s}\ket{\phi^{(k+2)}_{p,q}} = \fr{ S^{(k+2)}_{(r,s),(p,q)}}{S^{(k+2)}_{(1,1),(p,q)}} \ket{\phi^{(k+2)}_{p,q}},
\end{equation}
where $S^{(k+2)}_{(r,s),(p,q)}$ is the modular $S$-matrix.
This non-invertible symmetry plays a crucial role in the analysis of the Zamolodchikov RG flow \cite{Gaiotto2012} (for recent progress see \cite{Tanaka2024,Ambrosino2025}).
One characteristic property is that $L^{(k+2)}_{r,1}$ commutes with $\phi^{(k+2)}_{1,3}$.
Since the Zamolodchikov RG flow arises from a relevant deformation by the $\phi^{(k+2)}_{1,3}$ operator,
this implies that the non-invertible symmetry generated by $\{ L^{(k+2)}_{r,1} \}$ remains preserved along the RG flow.
This property manifests in the RG brane as the ability of the Verlinde lines $\{ L^{(k+2)}_{r,1} \}$ to pass through it topologically.
In the folded picture, the corresponding topological line can topologically end on the RG brane.

Consider an open Verlinde line crossing the Gaiotto RG brane, ending on $\phi^{(k+1)}_{1,r}$ and $\overline{\phi^{(k+2)}_{r,1}}$.
Through the folding trick, the non-local operator $\phi^{(k+1)}_{1,r} \, \phi^{(k+2)}_{r,1}$ is obtained,
\begin{equation}
\parbox{\compw}{\usebox{\boxcomp}}.
\end{equation}
Note that in the folded picture, the holomorphic and anti-holomorphic parts of $\mathrm{CFT}_R$ are interchanged.
In the representation of $\mathrm{CFT}_L \otimes \overline{\mathrm{CFT}}_R$, this operator corresponds to a chiral primary with spin
\begin{equation}
h=\fr{(r-1)^2}{2}.
\end{equation}
In particular, $\phi^{(k+1)}_{1,3} \, \phi^{(k+2)}_{3,1}$ is the spin-2 current, i.e. the {\it phantom current}.
The presence of the phantom current enables the interface to mix the stress tensor and the phantom current through rotations,  
which in turn enriches the possible structures of conformal interfaces.
In \cite{Gaiotto2012}, the rotation matrix is identified by reexpressing the representation of $\mathcal{M}(k+2,k+1) \otimes \mathcal{M}(k+3,k+2)$  
in terms of the representation of $\mathcal{SM}(k+3,k+1) \otimes \mathcal{M}(4,3)$,
which leads to the following relation for the stress tensor in the folded picture:
\begin{equation}
\begin{aligned}
T^{(k+1)} &= \fr{k+3}{2k+4} T_{\ca{SM}}
+ \frac{\sqrt{(k+1)(k+3)}}{2k+4} G\psi
+ \frac{k-1}{2k+4} T_{\psi}, \\
T^{(k+2)} &= \frac{k+1}{2k+4} T_{\ca{SM}}
- \frac{\sqrt{(k+1)(k+3)}}{2k+4} G\psi
+ \frac{k+5}{2k+4} T_{\psi}, \\
\phi^{(k+1)}_{1,3} &\phi^{(k+2)}_{3,1}
= (k+1)(k+3) T_{\ca{SM}}
- 3\sqrt{(k+1)(k+3)} G\psi \\
&- 3(k-1)(k+5) T_{\psi},
\end{aligned}
\end{equation}
where $T_{\mathcal{SM}}$ and $T_\psi$ are the stress tensors of $\mathcal{SM}(k+3,k+1)$ and $\mathcal{M}(4,3)$, respectively,  
and $G$ and $\psi$ correspond to the free fermion current and the superconformal generator.
In the representation of $\mathcal{SM}(k+3,k+1) \otimes \mathcal{M}(4,3)$, the UV–IR map reduces to a simple transformation: $\psi \to -\psi$.
By translating this back into the representation of $\mathcal{M}(k+2,k+1) \otimes \mathcal{M}(k+3,k+2)$,  
we finally obtain the following UV–IR map in the unfolded picture:
\begin{equation}
\begin{aligned}
  \left(
    \begin{array}{c}
     T^{(k+1)}    \\
     \overline{T^{(k+2)}}   \\
     \phi^{(k+1)}_{1,3} \overline{\phi^{(k+2)}_{3,1}}   \\
    \end{array}
  \right)
=\Lambda
  \left(
    \begin{array}{c}
     \overline{T^{(k+1)}}    \\
     T^{(k+2)}  \\
     \overline{\phi^{(k+1)}_{1,3}} {\phi^{(k+2)}_{3,1}}   \\
    \end{array}
  \right),
\end{aligned}
\end{equation}
where the explicit form of the matrix $\Lambda$ is given by
\begin{equation}\label{eq:Lambda}
\begin{aligned}
\Lambda
=
  \left(
    \begin{array}{ccc}
    \fr{3}{(k+2)(k+4)}   & \fr{(k-1)(k+3)}{k(k+2)} & \fr{1}{k(k+2)(k+4)} \\
    \fr{(k+1)(k+5)}{(k+2)(k+4)}   & \fr{3}{k(k+2)} & -\fr{1}{k(k+2)(k+4)} \\
     \fr{6(k+1)(k+5)}{k+4}  &  -\fr{6(k-1)(k+3)}{k}   &  1-\fr{6}{k(k+4)} \\
    \end{array}
  \right).
\end{aligned}
\end{equation}
One can extract $c_{LR}$ from $[\Lambda]_{12}$ as
\begin{equation}\label{eq:cLR}
c_{LR} = [\Lambda]_{12} \, c^{(k+2)}= \fr{(k-1)(k+5)}{(k+2)^2}.
\end{equation}
This is consistent with the perturbative result given in \cite{Brunner2015}.

\section{Transmission Coefficients from Phantom Currents}
\label{sec_coeff}

Building on the analysis of the Gaiotto RG brane, we construct a framework that can be applied to more general theories.
Specifically, motivated by the example of minimal models, we focus on cases in which the folded theory contains only two spin-2 currents -- the stress tensor and a phantom current.
This is for simplicity.
In fact, we expect a similar mechanism to hold even when multiple phantom currents are present.
However, since the number of equations to be solved increases and the analysis becomes cumbersome, we do not consider such cases in this Letter.
Our goal is to identify the minimal data required to determine the transmission coefficient, 
and to clarify the mechanism by which this determination is achieved.

In the following, we refer to spin-2 currents other than the ordinary currents with conformal weights $(h_L, h_R) = (2, 0) \text{ or } (0,2)$ as phantom currents.
We assume that the theory contains a single phantom current of the form $X = \phi_L \, \overline{\phi_R}$,  
where $\phi_L$ and $\overline{\phi_R}$ are primary operators with conformal dimensions $h_L$ and $h_R$, respectively,  
such that $h_L + h_R = 2$.
As seen in the example of the Gaiotto RG brane, when the folded theory possesses a phantom current,  
the conformal interface is allowed to have a structure that mixes the stress tensor and the phantom current in the following way:
\begin{equation}
\label{eq:gruing_spin2}
\begin{aligned}
  \left(
    \begin{array}{c}
     T_L    \\
     \overline{T}_R   \\
     X   \\
    \end{array}
  \right)
=\Lambda
  \left(
    \begin{array}{c}
     \overline{T}_L    \\
     T_R  \\
     \overline{X}   \\
    \end{array}
  \right).
\end{aligned}
\end{equation}
For convenience, we introduce undetermined coefficients $\alpha, \beta, \gamma, \alpha', \beta', \gamma'$ as follows:
\begin{equation}
\begin{aligned}
T_L&=\alpha \overline{T}_L + \beta T_R + \gamma \overline{X}, \\
\overline{T}_R&=\alpha' \overline{T}_L + \beta' T_R + \gamma' \overline{X}. \\
\end{aligned}
\end{equation}
We demonstrate that the values of these coefficients are fixed by the following constraints:
\begin{itemize}
  \item[(i)] \textbf{Preserving diagonal Virasoro symmetry}:
  \begin{equation}
  \label{eq:conformal_interface}
    T_L + \overline{T}_R = \overline{T}_L + T_R,
  \end{equation}
  which leads to the constraints
  \begin{equation}
    \alpha + \alpha' = 1, \quad 
    \beta + \beta' = 1, \quad 
    \gamma + \gamma' = 0.
  \end{equation}

  \item[(ii)] \textbf{Canonical normalization}:
\begin{equation}\label{eq:TT}
\begin{aligned}
    \braket{T_L(z)T_L(w)} &= \frac{c_L}{2(z-w)^4}+\cdots, \\
    \braket{\overline{T}_R(z)\overline{T}_R(w)} &= \frac{c_R}{2(z-w)^4}+\cdots, \\
    \braket{X(z)X(w)} &= \frac{N_X}{(z-w)^4}+\cdots,
\end{aligned}
\end{equation}
where $c_L$ ($c_R$) is the central charge of the CFT on the left (right) side of the interface.
These imply the following relations:
\begin{align}
    \alpha^2 c_L + \beta^2 c_R + 2 \gamma^2 N_X &= c_L, \label{eq:norma}\\
    \alpha'^2 c_L + \beta'^2 c_R + 2 \gamma'^2 N_X &= c_R. \label{eq:normb}
\end{align}
The normalization $N_X>0$ is arbitrary, since the transmission coefficients are independent of $N_X$.

  \item[(iii)] \textbf{Cluster decomposition principle}:
  \begin{equation}
    \braket{T_L(z)\, \overline{T}_R(w)} = 0.
  \end{equation}
  From the coefficient of the fourth-order pole, we obtain
\begin{equation}\label{eq:clua}
    \alpha\alpha' c_L + \beta\beta' c_R + 2 \gamma\gamma' N_X = 0.
\end{equation}
Since this is obtained as a sum (\ref{eq:norma}) + (\ref{eq:normb}), it provides no additional constraint.
From the coefficient of the second-order pole, we obtain
\begin{align}
\alpha\alpha'+\gamma \gamma' N_X \fr{h_L}{c_L}&=0, \label{eq:club}  \\
\beta\beta'+\gamma \gamma' N_X \fr{h_R}{c_R}&=0.  \label{eq:cluc}
\end{align}

\end{itemize}
From equations (\ref{eq:clua}), (\ref{eq:club}), and (\ref{eq:cluc}), we obtain
\begin{equation}
(2 - h_L - h_R) \, N_X \, \gamma \gamma' = 0.
\end{equation}
This implies that either $h_L + h_R = 2$ or $\gamma = 0$.
This is consistent with the fact that the conformal interface allows mixing between only spin-2 currents.
In the end, among equations (\ref{eq:norma}), (\ref{eq:normb}), (\ref{eq:club}), and (\ref{eq:cluc}),  
only three are independent, and their solution is given by:

\noindent
\textbf{Case (a) $\gamma=0$ and $c_L\neq c_R$:}

In this case, (\ref{eq:club}) and (\ref{eq:cluc}) imply
\begin{equation}
\alpha\alpha'=0,\qquad \beta\beta'=0.
\end{equation}
Thus, the solution to (\ref{eq:norma}) is given by
\begin{equation}
\alpha=1, \ \ \ \beta=0, \ \ \ \gamma=0.
\end{equation}
This corresponds to a factorized interface (i.e. a totally reflective interface) \cite{Popov2025},
which satisfies $c_{LR}=0$.
It indicates that when there is no spin-2 current other than the stress tensor,  
no nontrivial interface can be constructed if $c_L\neq c_R$.

\noindent
\textbf{Case (b) $\gamma=0$ and $c_L = c_R$:}

In this case, beyond the solution found in Case (a), the following additional solution is admissible:
\begin{equation}
\alpha=0, \ \ \ \beta=1, \ \ \ \gamma=0.
\end{equation}
This solution corresponds to a topological interface (i.e. a totally transmissive interface),
which satisfies $c_{LR}=c_L=c_R$.

\noindent
\textbf{Case (c) $\gamma\neq0$ and $c_L h_L \neq c_R h_R$:}

In this case, the solution is given by
\begin{equation}\label{eq:casec}
\begin{aligned}
\alpha &= \frac{h_L(c_R-c_L)}{c_R h_R - c_L h_L},  \\
\beta  &= \frac{c_L(h_R-h_L)}{c_R h_R - c_L h_L},  \\
\gamma^2 &= \frac{c_Lc_R(c_R-c_L)(h_R-h_L)}{N_X(c_R h_R - c_L h_L)^2}.
\end{aligned}
\end{equation}
This will later be shown to correspond to the Gaiotto RG brane.
The coefficient $c_{LR}$ is given by
\begin{equation}
c_{LR}=\beta c_R = (1-\alpha)c_L = \frac{c_L c_R(h_R-h_L)}{c_R h_R - c_L h_L}.
\end{equation}
This observation partially addresses the question of what specific data in the theory determines the transmission coefficient.  
In the case where a single phantom current is present, the transmission coefficient is entirely determined by its conformal dimension.

\noindent
\textbf{Case (d) $\gamma\neq0$ and $c_L h_L = c_R h_R$:}

In this case, the system of equations loses rank.
From equations (\ref{eq:norma}), (\ref{eq:normb}), (\ref{eq:club}), and (\ref{eq:cluc}), we obtain
\begin{equation}\label{eq:crcl}
(1-\alpha)(c_L-c_R)=0.
\end{equation}
If $c_L \neq c_R$, the only allowed solution is the factorized interface,
\begin{equation}
\alpha=1, \ \ \ \beta=0, \ \ \ \gamma=0.
\end{equation}
as in Case (a).
If $c_L = c_R$,
one finds a continuous one-parameter family of solutions
\begin{equation}\label{eq:sold}
\alpha+\beta=1,\qquad 
\gamma^2=\frac{c_L}{N_X}\,\alpha(1-\alpha),
\end{equation}
which can be parametrized as
\begin{equation}\label{eq:cased}
\begin{aligned}
\alpha&=\cos^2 2\theta, \\
\beta&=\sin^2 2\theta,  \\
\gamma&=\pm \sqrt{\frac{c_L}{4N_X}}\,\sin 4\theta.
\end{aligned}
\end{equation}
As we will see later, this corresponds to the interface in the free boson theory.  
The free boson admits a continuous family of theories connected by exactly marginal deformations,  
and accordingly, the conformal interfaces are parametrized by continuous parameters.

\section{Examples}

\noindent
\textbf{Minimal models:}

As input data, we consider the following set of quantities for the Gaiotto RG brane.
\begin{equation}
\begin{aligned}
c_L&=1-\fr{6}{(k+1)(k+2)}, \ \ \ c_R=1-\fr{6}{(k+2)(k+3)},\\
h_L&=\fr{k}{k+2}, \ \ \ \ \ \ \ \ \ \ \ \ \ \ \ \ \ \ \    h_R=\fr{k+4}{k+2}.
\end{aligned}
\end{equation}
With this set, our result (\ref{eq:casec}) remarkably reproduces the corresponding UV-IR map (\ref{eq:Lambda}),
which provides a justification for our phantom current mechanism.

As a nontrivial example, let us consider the brane corresponding to the RG flow from $\mathcal{M}(2q+1,q)$ to $\mathcal{M}(q,2q-1)$ \cite{Martins1992,Martins1992a}.
At present, the explicit construction of this RG brane is not known, and consequently, the UV-IR map for spin-2 currents has not been computed.
In fact, the folded theory $\mathcal{M}(q,2q-1) \otimes \mathcal{M}(2q+1,q)$ contains a spin-2 phantom current of the form $\phi^{(q,2q-1)}_{1,2} \, \phi^{(2q+1,q)}_{2,1}$. 
Therefore, our method allows us to construct the UV-IR map for the spin-2 currents.
Specifically, by substituting the following input data into equation~(\ref{eq:casec}), 
\begin{equation}
\begin{aligned}
c_L &= 1 - \frac{6 (q-1)^2}{q \, (2q -1)}, \quad
c_R = 1 - \frac{6 (q+1)^2}{q \, (2q + 1)}, \\
h_L &= \fr{4q-3}{4q},  \ \ \ \ \ \ \ \ \ \ \ \ 
h_R = \fr{4q+3}{4q},
\end{aligned}
\end{equation}
we obtain
\begin{equation}
\begin{aligned}
\alpha&=\frac{2 \, (4 q -3)\, (3 q^2 -1)}{q \, (32 q^2 - 7)}, \\
\beta&=\frac{(q - 2)\, (2 q + 1)\, (4 q - 3)}{q \, (32 q^2 - 7)}, \\
\gamma^2&=-\frac{8 \, (q -2)\, (q + 2)\, (4 q - 3)\, (4 q + 3)\, (3 q^2 - 1)}{q^2 \, (32 q^2 - 7)^2}.
\end{aligned}
\end{equation}
Furthermore, the coefficient $c_{LR}$ can be computed from this result as
\begin{equation}
c_{LR}=-\frac{(q - 2)\,(q + 2)\,(4 q - 3)\,(4 q + 3)}{q^2 \, (32 q^2 - 7)}.
\end{equation}
Since this is a non-unitary theory, we are currently unable to provide a precise interpretation of the result. 
Nevertheless, we hope that this result may serve as a guiding clue toward comparison with numerical simulations in non-unitary CFTs.
In fact, given the current lack of understanding of transmission coefficients in non-unitary theories, 
our general framework may offer a significant stepping stone toward further insight.
It should be noted, however, that our results rely on the cluster decomposition principle,
which may suggest that applying our method to (certain) non-unitary CFTs requires some care or modification.

\noindent
\textbf{Free bosons:}

We consider a compact free boson CFT with field identification $\phi \sim \phi + 2\pi R$,
whose action is given by
\begin{equation}
S=\frac{1}{8\pi}\int d^2 z\, \partial\phi\,\bar\partial\phi.
\end{equation}
The holomorphic/anti-holomorphic stress tensors are
\begin{equation}
T=-\tfrac{1}{2}:\!\partial\phi\,\partial\phi\!:,\qquad
\overline{T}=-\tfrac{1}{2}:\!\bar\partial\phi\,\bar\partial\phi\!:.
\end{equation}
This theory admits an exactly marginal deformation generated by $-\partial\phi\,\bar\partial\phi$,
which changes the compactification radius as $R\to R'\equiv R\ex{\pi \lambda}$.
In the folded theory consisting of free bosons before and after the deformation, there exists a local spin-2 phantom current constructed from the $\mathfrak{u}(1)$ current $i\partial\phi$, given by $X = -\partial \phi_L \, \bar{\partial} \phi_R$.
The interface connecting the undeformed and deformed theories has been constructed in \cite{Bachas2002, Bachas2012}, which glues the left/right currents as
\begin{equation}
  \left(
    \begin{array}{c}
       \del \phi_L   \\
       \bar{\del}\phi_R  \\
    \end{array}
  \right)
  =
  S_\pm 
    \left(
    \begin{array}{c}
       \bar{\del}\phi_L   \\
       \del \phi_R  \\
    \end{array}
  \right),
\end{equation}
where $S_\pm$ is the gluing matrix parameterized by $\theta$ with $\tan \theta = \mathrm{e}^{\pi \lambda}$ as
\begin{equation}
  S_\pm = 
    -\left(
    \begin{array}{cc}
     \cos 2\theta  & \pm \sin 2\theta   \\
      \sin 2\theta & \mp \cos 2\theta  \\
    \end{array}
  \right).
\end{equation}
From this, we find that the rotation matrix for the spin-2 currents takes the following form
\begin{equation}\label{eq:freeL}
\begin{aligned}
\Lambda_\pm
=
  \left(
    \begin{array}{ccc}
	\cos^2 2\theta & \sin^2 2 \theta & \pm \fr{\sin 4 \theta}{2} \\
	\sin^2 2\theta & \cos^2 2 \theta & \mp \fr{\sin 4 \theta}{2} \\
	\sin 4 \theta  & -\sin 4 \theta &\mp \cos 4 \theta.
    \end{array}
  \right).
\end{aligned}
\end{equation}
It follows that the transmission coefficient is $c_{LR} = \sin^2 2\theta$, 
in agreement with \cite{Quella2007}.

We now demonstrate that this is reproduced by our phantom current mechanism.
The required input data are as follows:
\begin{equation}
c_L=c_R=1, \ \ \ h_L=h_R=1.
\end{equation}
This corresponds to Case (d) in our classification of the solutions.
We have shown that the solution in this case is given by a continuous family, as in (\ref{eq:cased}),
which exactly matches the gluing matrix in (\ref{eq:freeL}).
This confirms that our mechanism correctly captures not only the Gaiotto RG brane but also a broader class of interfaces.

\section{Discussions}

In this work, we have developed a general framework for determining the transmission coefficient across a conformal interface.
This framework encompasses a broad class of models, including prominent examples such as minimal models and the free boson CFT.
However, in this letter, to highlight the essence of the proof, we remove inessential complications by assuming that there are only two spin-2 currents, {\it i.e.} a unique stress tensor and a single phantom current.
In more general cases, one has to solve more equations, yet we expect the same mechanism to go through.
Another possibility not captured by our mechanism is the case where the spin-2 phantom current does not factorize into two primaries — for example, the Oshikawa–Affleck defect \cite{Oshikawa1997}.
In fact, this can still be determined by almost the same mechanism, as discussed in Appendix \ref{app:one}.
Extending the classification to a broader class is an important direction.

While we have focused in particular on the transmission coefficient,
the framework presented here is expected to be applicable also to the determination of other quantities, such as the effective central charge, $c_{\mathrm{eff}}$ \cite{Sakai2008, Brehm2015, Wen2018, Karch2023,Karch2024, Barad2025}.
Furthermore, the mechanism obtained in this work yields a universal fusion rule for the transmission coefficient,
which may provide a new approach to understanding the fusion of non-topological interfaces \cite{Bachas2007,Kravchuk2024,Diatlyk2024}.
We hope to report the details, together with justification based on numerical calculations, in a future paper.

We introduced the concept of the spin-2 phantom current and discussed its role. We expect that the benefits of this new concept will extend beyond merely determining the transmission coefficient.
A particularly intriguing direction is to develop a systematic algebraic framework for the spin-2 phantom current and apply it to the explicit construction of new interfaces, which we hope to report on in the future.

\section*{Acknowledgments}
We would like to thank Yifan Wang for helpful comments on the draft. We thank Yu-ki Suzuki for collaboration at an early stage of this work. We also thank Yichul Choi, Yifan Wang, and Yunqin Zheng for valuable discussions on phantom currents.
YK is supported by the INAMORI Frontier Program at Kyushu University and JSPS KAKENHI Grant Number 23K20046.

\bibliographystyle{JHEP}
\bibliography{main.bib}

\clearpage
\onecolumngrid
\appendix

\section{Spin-$1$ Phantom Current}\label{app:one}
In this section, we consider the situation in which the folded theory has a single spin-1 phantom current $j(z)=\psi_L(z)\overline{\psi}_R(z)$ (and its anti-holomorphic counterpart). The simplest example of such a theory is the Ising CFT, which has three primary operators, $\bb{I}$, $\sigma$, and $\epsilon$, and three Verlinde lines, $1$, $\eta$, and $\ca{N}$.
The line $\eta$ corresponds to the anomaly-free $\mathbb{Z}_2$ symmetry, and the twisted Hilbert space $\mathcal{H}_\eta$ consists of three primaries:
\begin{equation}
    \psi_{\fr{1}{2},0},\quad \overline{\psi}_{0, \fr{1}{2}}, \quad \mu_{\fr{1}{16}, \fr{1}{16}.}
\end{equation}
Here, $\psi$ and $\overline{\psi}$ are left and right moving Majorana fermion. 
In the folded theory, the diagonal $\mathbb{Z}_2$ twisted Hilbert space possesses the following spin-1 currents:
\begin{equation}
    j= \psi_L \overline{\psi}_{R},\quad \overline{j}=\overline{\psi}_L \psi_R.
\end{equation}
In this case, a phantom symmetry
\begin{equation}
    U_\epsilon= e^{i\epsilon \int \fr{dz}{2\pi i}j(z)} +  e^{-i\epsilon \int \fr{dz}{2\pi i}j(z)} 
\end{equation}
exists, allowing deformations such that the boundary condition \eqref{eq:conformal_interface} holds on the defect for the topological defects $1$, $\eta$, and $\mathcal{N}$, yielding a defect conformal manifold~\cite{Antinucci2025}.

The key point in our discussion is the existence of the spin-2 operator. The operator
\begin{equation}
    W^+ \propto c_RT_L -c_L\overline{T}_R
\end{equation}
is a primary operator of conformal dimension $(2,0)$ with respect to $\mathrm{Vir} \times \overline{\mathrm{Vir}}$ in $\mathrm{CFT}_L \times \overline{\mathrm{CFT}_R}$,
which determines the transmission coefficient~\cite{Quella2007}. 
When a spin-1 current exists, we can construct an $\eta$-twisted $(2,0)$ primary field in addition to this operator~\cite{Antinucci2025}
\begin{equation}
    W^- \propto h_R \partial \psi_L \overline{\psi}_R - h_L\psi_L\overline{\partial\psi}_R.
\end{equation}
This $W^-$ operator is clearly linearly independent from $T_L$ and $T_R$, and on non-trivial interfaces, it mixes with them via eq~\eqref{eq:gruing_spin2}
\begin{equation}
\label{eq:gluing_w_eq}
\begin{aligned}
  \left(
    \begin{array}{c}
     T_L    \\
     \overline{T}_R   \\
     W^-   \\
    \end{array}
  \right)
=\Lambda
  \left(
    \begin{array}{c}
     \overline{T}_L    \\
     T_R  \\
     \overline{W}^-   \\
    \end{array}
  \right).
\end{aligned}
\end{equation}

Let us consider the defect conformal manifold obtained from the defect exactly marginal deformation in the Ising CFT.
The defects in this one-parameter family can be expressed in terms of the deformation parameter $\gamma$ as the following gluing condition for the fermions~\cite{Oshikawa1996,Oshikawa1997,Bachas2013}:
\begin{equation}
  \left(
    \begin{array}{c}
       \psi_L   \\
        -i\overline{\psi}_L  \\
    \end{array}
  \right)
  =
  S_\pm 
    \left(
    \begin{array}{c}
       \psi_R   \\
        -i\overline{\psi}_R  \\
    \end{array}
  \right),
\end{equation}
\begin{equation}
  S_\pm = 
    -\left(
    \begin{array}{cc}
     \cosh \gamma  & \pm \sinh \gamma   \\
      \sinh \gamma & \pm \cosh \gamma  \\
    \end{array}
  \right).
\end{equation}
The boundary condition in the folded theory derived from these gluing conditions is
\begin{equation}
    \left[ \begin{pmatrix} \psi^L_r \\ \psi^R_r \end{pmatrix} + i\mathcal{O} \begin{pmatrix} \overline{\psi}^L_{-r} \\ \overline{\psi}^R_{-r} \end{pmatrix} \right] |B\rangle\rangle = 0,
\end{equation}
where 
\begin{equation}
    \mathcal{O}_{\pm} = \begin{pmatrix} \cos(2\theta) & \sin(2\theta) \\ \pm\sin(2\theta) & \mp\cos(2\theta) \end{pmatrix}, 
\end{equation}
\begin{equation}
    \cos(2\theta) = \tanh\gamma \iff e^{\gamma} = \cot\theta .
\end{equation}
We can explicitly calculate the mixing (\ref{eq:gluing_w_eq}) from the fermionic gluing conditions and choosing the normalization $W^-=\fr{1}{2}( \partial \psi_L \overline{\psi}_R -\psi_L\overline{\partial\psi}_R$). The explicit form of the matrix $\Lambda$ is found to be:
\begin{equation}\label{eq:lambda_ising}
\Lambda_{\pm} =
\begin{pmatrix}
\cos^2(2\theta) & \sin^2(2\theta) & -\frac{1}{2}\sin(4\theta) \\
\sin^2(2\theta) & \cos^2(2\theta) & \pm\frac{1}{2}\sin(4\theta) \\
\mp\sin(4\theta) & \pm\sin(4\theta) & \mp\cos(4\theta)
\end{pmatrix}.
\end{equation}

As in Section~\ref{sec_coeff}, the part of $\Lambda$ can also be calculated from \eqref{eq:TT} and \eqref{eq:clua}. Instead of \eqref{eq:club} and \eqref{eq:cluc}, we obtain
\begin{align}
    2\alpha\alpha' + \gamma\gamma'c_{W^-W^-T_L}&=0, \label{eq:clud}\\
    2\beta\beta' + \gamma\gamma'c_{W^-W^-\overline{T}_R}&=0. \label{eq:clue}
\end{align}
The OPE coefficients of the spin-2 currents are 
\begin{align}
    c_{W^-W^-T_L} &= -\fr{4N_{W^-}h_Lh_R}{c_L}(h_L^2 + h_L h_R + h_R) \label{eq:coeff_l},\\
    c_{W^-W^-\overline{T}_R} &= -\fr{4N_{W^-}h_Lh_R}{c_R}(h_R^2 + h_L h_R + h_L) \label{eq:coeff_r},
\end{align}
where $N_{W^-}$ is the normalization factor of $W^-$.
Therefore, from \eqref{eq:norma}, \eqref{eq:normb}, \eqref{eq:clud}, and \eqref{eq:clue},  we obtain the equation \eqref{eq:crcl} under the condition $c_L(h_L^2 + h_Lh_R + h_R) = c_R(h_R^2 + h_Lh_R + h_L)$ instead of the condition $c_Lh_L=c_R h_R$. Since $h_L+ h_R=1$, this condition becomes $c_L = c_R$. Thus, in a similar way to (\ref{eq:sold}), we obtain $\alpha + \beta = 1$ and $\gamma^2 = \frac{2\alpha(1-\alpha)}{c_{W^- W^- T_L}}$, which reproduce the transmittance calculated in \eqref{eq:lambda_ising}.

\end{document}